\documentclass[aps,prc,twocolumn,times,graphicx,tighten,showpacs]{revtex4}
\usepackage[dvips]{graphicx}

\newcommand{\bea}{\begin{eqnarray}}
\newcommand{\eea}{\end{eqnarray}}
\newcommand{\be}{\begin{equation}}
\newcommand{\ee}{\end{equation}}
\newcommand{\ba}{\begin{eqnarray}}
\newcommand{\ea}{\end{eqnarray}}

\begin{document}

\title{Scaling and isospin effects in quasielastic lepton-nucleus scattering in
the Relativistic Mean Field Approach
}

\author{
J.A. Caballero$^a$,
J.E. Amaro$^b$,
M.B. Barbaro$^c$
T.W. Donnelly$^d$,
J.M. Ud\'{\i}as$^e$
}
\affiliation{$^a$Departamento de F\'{\i}sica At\'omica, Molecular y Nuclear,
Universidad de Sevilla, 41080 Sevilla, SPAIN}
\affiliation{$^b$Departamento de F\'{\i}sica At\'omica, Molecular y Nuclear,
Universidad de Granada, 18071 Granada, SPAIN}
\affiliation{$^c$Dipartimento di Fisica Teorica, Universit\`a di Torino and
  INFN, Sezione di Torino, Via P. Giuria 1, 10125 Torino, ITALY}
\affiliation{$^d$Center for Theoretical Physics, Laboratory for Nuclear
  Science and Department of Physics, Massachusetts Institute of Technology,
  Cambridge, MA 02139, USA}
\affiliation{$^e$Departamento de F\'{\i}sica At\'omica, Molecular y Nuclear,
  Universidad Complutense de Madrid, 28040 Madrid, SPAIN}

\date{\today}


\begin{abstract}
  The role of isospin in quasielastic electron scattering and charge-changing neutrino
  reactions is investigated in the relativistic
  impulse approximation. We analyze proton and neutron scaling functions
  making use of various theoretical descriptions for the final-state
  interactions, focusing on the effects introduced by the presence of strong
  scalar and vector terms in the relativistic mean field approach.  An
  explanation for the differences observed in the scaling functions evaluated
  from $(e,e')$ and $(\nu,\mu)$ reactions is provided by invoking the
  differences in isoscalar and isovector contributions.
\end{abstract}


\maketitle

Extensive analyses of quasielastic (QE) inclusive electron
scattering data performed in recent
years~\cite{Day90,DS199,DS299,MDS02} have clearly demonstrated the
quality of the behavior known as scaling. These analyses are based
on the so-called superscaling function, $f(\psi')$, obtained by
dividing the cross section by an appropriate function which contains
the single nucleon physics, and plotting the result against the
scaling variable $\psi'(q,\omega)$ (see, e.g.,
\cite{Barbaro:1998gu}). One then studies the dependences upon the
momentum transfer $q$ and the specific nucleus chosen. From
$(e,e')$ world data one concludes that scaling of the {\em first
kind} (no dependence on $q$) is reasonably respected at energies
below the QE peak, whereas scaling of the {\em second kind} (no
dependence on the nuclear species) is fulfilled very well in the
same region. The simultaneous occurrence of both kinds of scaling is
called {\em superscaling}. At energies above the QE peak, breaking
of both kinds of scaling is observed, residing mostly in the
transverse channel, and likely due to effects beyond the impulse
approximation.

Experimental data lead to a scaling function with a characteristic
asymmetric shape, having a tail that extends to high values of the
transferred energy $\omega$ (positive values of the scaling variable
$\psi'(q,\omega)$).  The asymmetric shape of the scaling function is
largely absent in non-relativistic (NR) models based on a mean field
approach. In contrast, the study presented in~\cite{PRL05,JAC06} has
shown that the asymmetry can in fact be obtained within the
relativistic impulse approximation (RIA), given that a description
of final-state interactions (FSI) using strong relativistic mean
field (RMF) potentials is assumed. Recently, we have
shown~\cite{DEB07} that an asymmetrical scaling function can be also
obtained within the framework of a semi-relativistic (SR) model,
based on improved NR expansions of the on-shell electromagnetic
current, provided that the FSI are described by the Dirac
equation-based (DEB) potential~\cite{Udias} derived from the RMF.
Note that, in the SR model,  the nonlocalities arising from the NR
reduction of the Dirac equation
--- and  incorporated into the wave functions through the Darwin
factor --- are essential to reproduce the asymmetric shape of the
scaling function.

The data analysis of the separated longitudinal ($L$) and transverse ($T$)
contributions to the scaling function carried out
in~\cite{DS299,MDS02} has shown that, whereas the $L$ response does
scale to a universal curve, the $T$ strength increases with the
transfer momentum $q$ and/or the mass number $A$.
This excess of strength in the transverse channel is not entirely
understood, although different effects ranging from FSI effects to
MEC contributions have been invoked to explain it. This result also
connects with the breaking of the {\em zeroth-kind} scaling (defined
as the equality of the scaling functions obtained from the separated
$L$ and $T$ contributions, viz., $f_L(\psi')=f_T(\psi')=f(\psi')$)
observed in the data. From previous
studies~\cite{JAC06,DEB07,Amaro05} it turned out that the
zeroth-kind scaling is closely fulfilled by various models based on
the impulse approximation. This is the case of the Relativistic
Fermi Gas (RFG), by construction. The use of traditional NR and SR
approaches also leads to similar longitudinal and transverse scaling
functions. This occurs for different descriptions of the FSI,
namely, using the same Woods-Saxon potential as in the initial
state, which leads to symmetrical scaling functions that do not
agree with experiment~\cite{Amaro05}, and making use of the DEB
potential that produces the correct amount of asymmetry in the
scaling function~\cite{DEB07}. Finally, the RIA also leads to the
fulfilment of zeroth-kind scaling when the plane wave limit is
assumed~\cite{JAC06}, i.e., the final nucleon state is described as
a free relativistic on-shell particle.  This is known as
Relativistic Plane Wave Impulse Approximation (RPWIA).

\begin{figure}[htb]
\begin{center}
\vspace{0.6cm}
\includegraphics[scale=0.47]{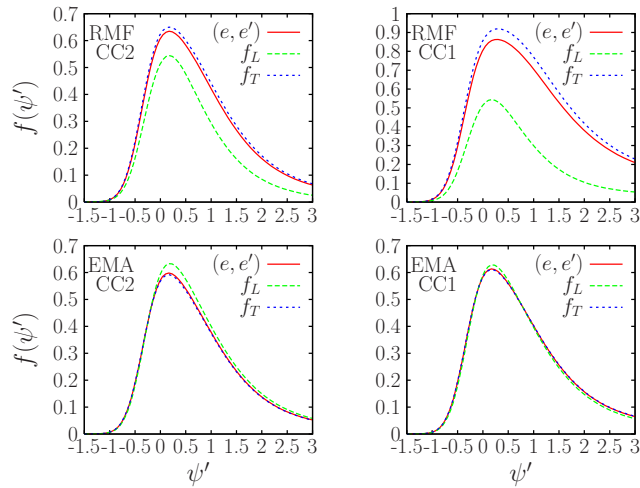}
\caption{(Color online) Scaling function $f(\psi')$ compared with
separate $L$ and $T$ contributions.  All results correspond to the
analysis of QE $(e,e')$ on $^{12}$C. For the separate scaling
functions the momentum transfer is $q=1$ GeV/c, while for the
$(e,e')$ results the incident electron energy is $\varepsilon=1$
GeV. The results correspond to the RMF description of FSI (top
panels) and the EMA approach (bottom). The current operators CC2
(left) and CC1 (right) have been considered.} \label{Fig1}
\end{center}
\end{figure}

Contrary to the above-mentioned models, breakdown of zeroth-kind
scaling is observed in the RIA model with FSI  described by the
strong scalar and vector potentials of the RMF, which yields a
scaling function found to be in excellent agreement with the data.
The amount of scaling violation depends on the particular
prescription chosen for the current operator.  This is illustrated
in the upper panels of Fig.~\ref{Fig1}, where we show $f(\psi')$ as
well as its longitudinal $f_L(\psi')$ and transverse $f_T(\psi')$
contributions for $^{12}$C. The momentum transfer has been fixed to
$q = 1$ GeV/c when computing $f_{L,T}$, while, for the global
scaling function $f$, in addition the beam energy has been fixed to
$\varepsilon = 1$ GeV, implying that the scattering angle is a
function of $\psi'$. The left and right panels correspond to results
for the CC2 and CC1 prescriptions of the electromagnetic current
operator, respectively \cite{Forest,JAC06}. In both cases breaking
of zeroth-kind scaling is clearly observed, $f_T$ being about $20\%$
larger than $f_L$ for CC2 and almost twice as large for CC1. This
result, first observed in~\cite{JAC06}, differs from the RPWIA and
SR models considered in the literature~\cite{JAC06,DEB07,Amaro05}
where zeroth-kind scaling is well obeyed. Moreover, it turns out in
the so-called scaling region ($\psi'<0$) that, within the RMF model,
scaling of the first kind is fulfilled separately by $f_L$ and $f_T$
with only modest scaling violations in each case. However, the two
scaling functions obtained are different, with $f_T$ lying higher
than $f_L$.

We should point out once more that, in contrast, the SR approach
with a convenient description of FSI based on the DEB potential
(derived from the same relativistic Hartree potential) yields
results~\cite{DEB07} consistent with the asymmetry shown by the RMF
model and data, but which respect zeroth-kind scaling almost
exactly. Although extensive tests showing the reliability of the SR
expansion~\cite{SRtest} have been performed within the context of
the RFG and RPWIA approaches, the effects introduced by the
dynamical enhancement of the lower components in the Dirac
spinors~\cite{NPA,Cris,Lund}, accounted for within a fully
relativistic calculation, are not present in the SR approaches and
this may be the reason that the fully relativistic and the SR
calculations for the scaling function differ.
From previous studies~\cite{PRL05,JAC06}, we have shown that the
scaling function is very slightly modified by the dynamical
relativistic effects linked to the initial bound nucleon states.

The presence of strong scalar and vector potentials in the nuclear
states (mainly the final one) leads to a significant enhancement of
the lower components of the four-spinors describing the nucleon wave
functions in the relativistic approach. Due to this, the ratio
between upper and lower components of the fully relativistic
calculation is quite different from the one for free spinors or
implied in the NR or SR approaches~\cite{Javi2004,Lund,Udias99}. In
order to assess the influence of the enhancement of the lower
components (also called spinor distortion \cite{Kelly}) the fully
relativistic calculation is compared here with the effective
momentum approach (EMA) \cite{Kelly,Lund,Udias,Javi2004}. In the EMA
the relationship between upper and lower components is forced to be
the same as for free spinors. Thus the EMA wave functions lack the
dynamical enhancement of the lower components due to the presence of
strong potentials.  This is also the case of SR approaches where
projections over positive-energy states and truncated expressions of
the current operator are considered.  Lacking this kind of
distortion, the EMA results should lie closer to the
so-called factorized result~\cite{Lund,Javi2004}.

Scaling of the zeroth kind is well fullfilled in the EMA model, as
is clearly illustrated in the bottom panels of Fig.~\ref{Fig1},
where we present the scaling functions again corresponding to the
two currents operators, now evaluated with the EMA. These results
illustrate the crucial role played by spinor distortion in leading
to zeroth-kind scaling violations. In fact, the EMA approach leads
to $f(\psi')\approx f_L(\psi')\approx f_T(\psi')$ using either of
the current operators.

This result is in accordance with the study performed in the context
of the SR approach and the DEB description of FSI~\cite{DEB07}. To
conclude, notice that projecting out the negative-energy components
yields an enhancement (decrease) of the $L$ ($T$) contributions. The
magnitude of these effects being moderate (strong) for CC2 (CC1)
current operators. This is in consonance with the known tendency of
the CC1 operator to enhance the effects of spinor distortion in the
electromagnetic observables, while CC2 shows a more moderate
dependence on the amount of distortion~\cite{NPA,Udias99,Javi2004}.

One important application of superscaling has been suggested in
\cite{SuSA} within the context of making realistic predictions for
charge-changing (CC) neutrino-nucleus differential cross sections
which, for example, are of interest in neutrino oscillation
experiments. The validity of the superscaling hypothesis, i.e., the
existence of a universal scaling function in electroweak processes,
constitutes an essential result which is supported by various
theoretical studies~\cite{PRL05,JAC06,DEB07,Amaro05} and gave rise
to recent applications to neutrino
studies~\cite{Martini:2007jw,Antonov:2006md}. The universal
character of the scaling function is the basis of the SuperScaling
Analysis (SuSA) introduced in~\cite{SuSA}. Within SuSA, the {\it
experimental} superscaling function extracted from the analysis of
$(e,e')$ world data is used to reconstruct CC neutrino-nucleus cross
sections.  However, it is important to point out that the extraction
of the experimental scaling function refers only to the analysis of
the longitudinal function, $f_L$, whereas, in contrast, $(\nu,\mu)$
reactions are totally dominated by the purely transverse
$T_{VV}+T_{AA}$ and $T'_{VA}$ channels. {
Thus, one may question the validity of using $f_L$
extracted from electron scattering data to predict $(\nu,\mu)$ cross
sections, which are dominated by transverse responses. This issue is
particularly relevant within theoretical frameworks which lead to
$f_L(\psi')\neq f_T(\psi')$, i.e., violation of zeroth-kind scaling.
In what follows we present a study of this issue within the RMF
approach. Our aim is to answer the above question and clarify the
degree to which the scaling hypothesis does or does not work.

\begin{figure}[htb]
\begin{center}
\includegraphics[scale=0.6]{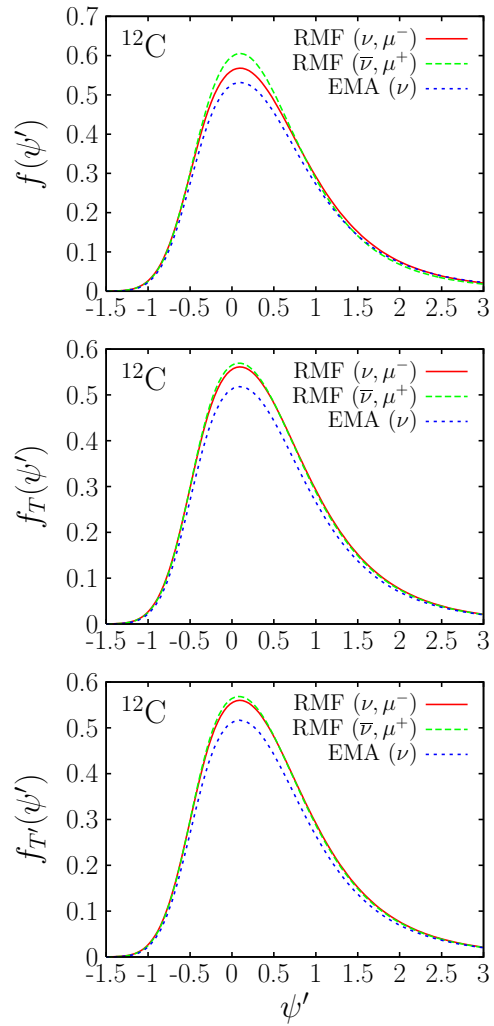}
\caption{(Color online) Scaling functions evaluated from
$(\nu,\mu^-)$ and $(\overline{\nu},\mu^+)$ reactions. The results
correspond to $\varepsilon_{\nu(\overline{\nu})}=1$ GeV and
$\theta_\mu=45^0$. Results from relativistic calculations using
the RMF are compared with EMA (see text for details). Top, middle
and bottom panels refer to the global scaling function, the
$T_{VV}+T_{AA}$ contribution and the $T'_{VA}$ one, respectively.}
\label{Fig2}
\end{center}
\end{figure}

In Fig.~\ref{Fig2} we present the scaling functions obtained from
the calculation of CC neutrino and antineutrino-nucleus reaction
cross sections. The kinematics correspond to neutrino (antineutrino)
energy fixed to $\varepsilon=1$ GeV and lepton scattering angle
$\theta_\mu=45^0$. In each graph we show the results obtained with
the RMF approach applied for neutrinos (solid line) and
antineutrinos (dashed) compared with EMA (in this case only the
curve for neutrinos is presented as the one for antineutrinos is
very similar). The separate analysis of the transverse channels, $T$
and $T'$, is also shown in middle and bottom panels, respectively.
The longitudinal contribution to inclusive CC neutrino-nucleus
scattering is negligible and therefore is not shown. The usual
relativistic single-nucleon expression for the charged-current
operator~\cite{Cris,SuSA} has been employed. From results in
Fig.~\ref{Fig2}, several basic conclusions emerge. First, the
scaling functions for both processes $(\nu,\mu^-)$ and
$(\overline{\nu},\mu^+)$ almost coincide. Second, it is verified
that $f(\psi')\approx f_T(\psi')\approx f_{T'}(\psi')$ to a high
degree, i.e., the separate transverse responses contribute similarly
to the global scaling function. Notice that this is valid, not only
for the EMA approach, but also within the fully RMF model. Third,
the dynamical enhancement of the lower components in the Dirac wave
functions leads only to a slight modification of the scaling
functions. These result are in accord with ones observed for
$(e,e')$ reactions in the CC2 case (see Fig.~\ref{Fig1}).

For the same kinematics, the basic difference between electron and
neutrino scattering is the nature of the exchanged vector boson, a
virtual photon probing the electromagnetic current in electron
scattering, and a $W^\pm$ probing the weak current in CC
neutrino-nucleus processes. As a consequence, whereas for inclusive
$(e,e')$ processes all nucleons (protons and neutrons) in the
nucleus contribute, in the case of CC neutrino (antineutrino)
reactions only the neutrons (protons) in the target contribute to
the inclusive cross section.

\begin{figure}[htb]
\begin{center}
\includegraphics[scale=0.6]{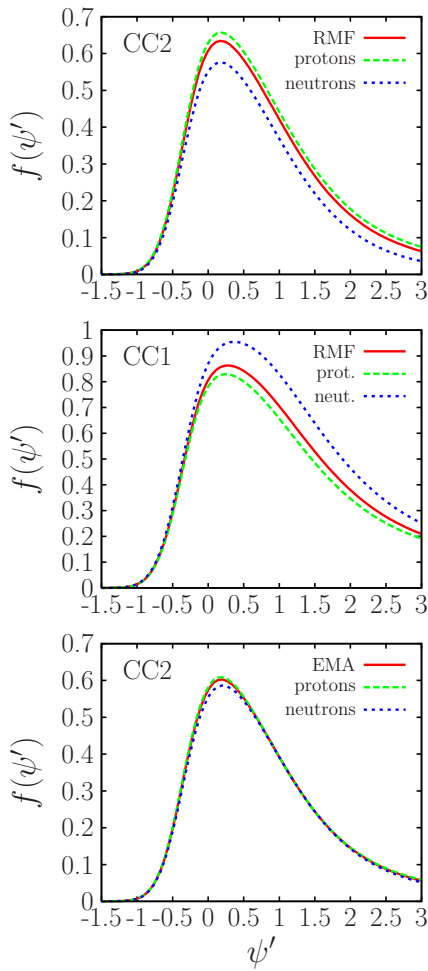}
\caption{(Color online) Proton and neutron contributions to the
scaling function obtained from the analysis of $^{12}$C$(e,e')$.
Same kinematics as in Fig.~1. The results correspond to the RMF
approach with CC2 (top panel) and CC1 (middle). The EMA case is
presented in the bottom panel.  } \label{Fig3}
\end{center}
\end{figure}

In order to compare $(e,e')$ and $(\nu,\mu^-)$ reactions, it is
convenient to separate the contributions of protons and neutrons
to the $(e,e')$ scaling function in the form
\begin{equation}
f(\psi') = \frac{G_P}{G}f_P(\psi') + \frac{G_N}{G} f_N(\psi') \, ,
\end{equation}
where $G_P$ and $G_N$ are the single-nucleon functions
(incorporating both the longitudinal $L$ and transverse $T$
contributions) defined in~\cite{MDS02,JAC06}
for protons and neutrons, respectively, and $G=G_P+G_N$.
The proton $f_P$ and neutron $f_N$ scaling functions are shown in
Fig.~\ref{Fig3} for the same kinematics as in Fig.~\ref{Fig1}. The
total scaling function, obtained as an average of $f_P$ and $f_N$
with weights $G_P/G$ and $G_N/G$, respectively, is also shown in the
figure with solid lines. We show results for the RMF approach and
for both current operators CC2 (top panel) and CC1 (middle). For
completeness, the results corresponding to EMA are also shown in the
bottom panel (here we only consider CC2, since results for CC1 are
basically the same). As one observes, in EMA the proton and neutron
scaling functions are almost identical i.e., $f_P=f_N$; such a
result will be called {\em scaling of the third kind} or isospin
scaling. A similar result also occurs for the NR and SR
calculations. In contrast, protons and neutrons yield different
scaling functions in the RMF approach, with magnitude dependent on
the choice of current operator. The results of Fig.~\ref{Fig3}
indicate that the proton/neutron balance in $f(\psi')$ is
significantly modified after the inclusion of dynamical relativistic
distortion of the spinors, due mainly to the presence of strong
relativistic potentials in the final state. Thus, the effect of
spinor distortion on the scaling function is isospin-dependent,
making the separate scaling functions for protons and neutrons
appreciably different, $f_P\ne f_N$, that is, violating scaling of
the third kind. This is connected to the breaking of the zeroth-kind
scaling observed in Fig.~\ref{Fig1} in the full relativistic
calculation. As will be shown later, the isoscalar/isovector terms
in both the $L$ and $T$ channels in $(e,e')$ reactions also yield
significant differences when compared with the purely isovector
contributions involved in CC neutrino-nucleus scattering processes.

\begin{figure}[htb]
\begin{center}
\includegraphics[scale=0.7]{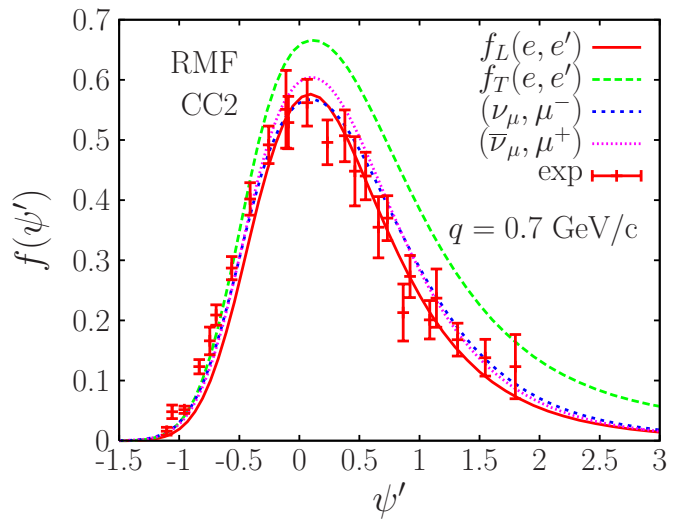}
\caption{(Color online) Longitudinal and transverse scaling
functions for $(e,e')$ compared with $f(\psi')$ evaluated from
$(\nu,\mu^-)$ and $(\overline{\nu},\mu^+)$. All results have been
evaluated with the RMF approach using the CC2 current operator.
The kinematics selected correspond to fixed values of the incident
lepton energy, $\varepsilon=1$ GeV, and momentum transfer, $q=0.7$
GeV/c. The averaged experimental function extracted from
longitudinal electron scattering data is also shown~\cite{MDS02}.}
\label{Fig4}
\end{center}
\end{figure}

The universal character of the scaling function and its validity for
electromagnetic and weak interactions is further analyzed in
Fig.~\ref{Fig4}. Here we directly compare the functions $f_L(\psi')$
and $f_T(\psi')$ obtained from $(e,e')$ cross sections with the ones
corresponding to $(\nu_\mu,\mu^-)$ and $(\overline{\nu}_\mu,\mu^+)$
reactions. All results have been evaluated within the RIA and making
use of the RMF potential to describe FSI. The prescription of the
current operator is CC2 for both electromagnetic and weak
interactions~\cite{JAC06}. The kinematics correspond to fixed values
of the incident lepton (electron, neutrino, antineutrino) energies,
$1$ GeV, and transferred momentum $q=0.7$ GeV/c. Similar results are
obtained if, instead of fixing the momentum transfer, we select a
specific value of the lepton (electron or muon) scattering angle.
The averaged QE phenomenological function obtained from the analysis
of $(e,e')$ data~\cite{MDS02} is also included in Fig.~\ref{Fig4}.
As observed, the theoretical curve for $f_L(\psi')$ follows the
behavior of the data very closely (referred only to the analysis of
the longitudinal scaling function), and this proves the capability
of the RIA combined with the RMF potential to describe $(e,e')$ data
in the longitudinal channel satisfactorily.  On the contrary, the
transverse contribution $f_T(\psi')$ overestimates the data by $\sim
20\%$ even in the region close to the maximum, $\psi'\approx 0$.
This result, arising from zeroth-kind scaling violation in the RMF
approach, is not in conflict with $(e,e')$ data that indeed leaves
room for effects of this type (see the general discussion
in~\cite{DS299,MDS02}).

Concerning the scaling function obtained for neutrino (and
antineutrino) scattering reactions, one observes that it is much
more in accordance with $f_L(\psi')$ and hence with the electron
scattering longitudinal data, than with $f_T(\psi')$. Within the
context of our model this outcome reinforces the validity of the
general assumption implied by SuSA~\cite{SuSA}, i.e., the use of the
phenomenological scaling function (extracted from the analysis of
longitudinal QE electron scattering data) to predict CC
neutrino-nucleus cross sections. However, it is also striking that
$f(\psi')$ for $\nu_\mu$ and $\overline{\nu}_\mu$ reactions, which
are totally dominated by the purely {\em transverse} ($T$, $T'$)
channels, coincides with the $f_L$ function of $(e,e')$ instead of
$f_T$, in contrast to what one might expect.

In order to understand these results, let us start by discussing
some basic differences between $(e,e')$ and $(\nu,\mu)$ reactions.
In the former, the longitudinal and transverse channels contribute
importantly (at least for some kinematics), and in both responses
isoscalar and isovector form factors enter. However, as shown in
Fig.~\ref{Fig3}, the balance between isovector and isoscalar
contributions may change in a significant way due to the strong
dynamical enhancement of the lower components in the outgoing
nucleon Dirac wave functions obtained with the RMF potential.  In
contrast, only purely isovector form factors enter in CC
neutrino-nucleus scattering. Hence, it is important to evaluate
isovector and isoscalar contributions in $(e,e')$ reactions and
their effects on the scaling functions. In what follows we
investigate how the functions $f_L$ and $f_T$ obtained from $(e,e')$
RMF calculations change when the isoscalar form factors are removed.
Notice that proceeding in this way, we force the $(e,e')$ to be
purely isovector, similar to what occurs for $(\nu,\mu)$.

\begin{figure}[htb]
\begin{center}
\includegraphics[scale=0.6]{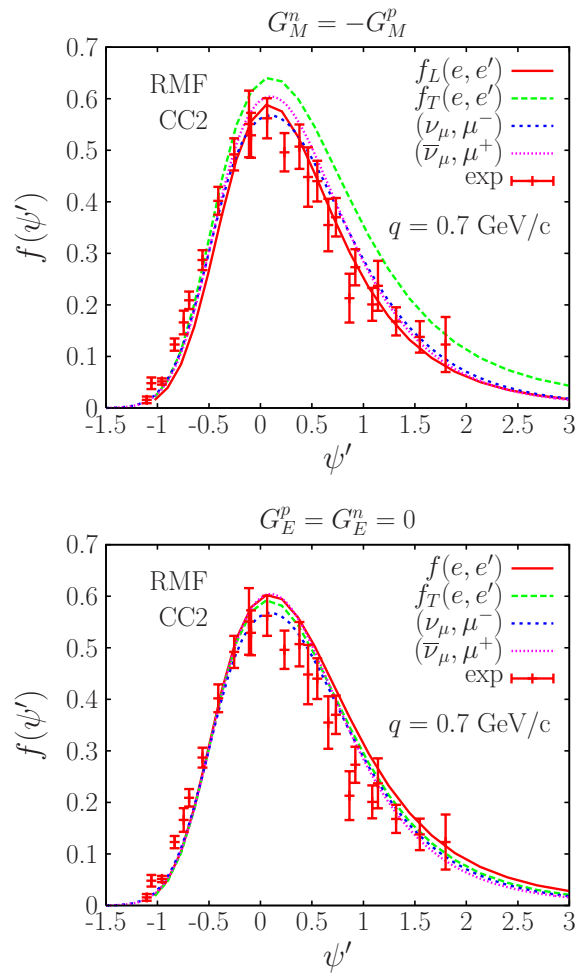}
\caption{(Color online) Same as Fig.~4, but with modified
isoscalar/isovector contributions via the nucleon form factors in
the $(e,e')$ case. Top panel: $G_M^n=-G_M^p$, i.e., magnetic form
factor purely isovector. Bottom panel: $G_E^n=G_E^p=0$, i.e., no
convective terms. } \label{Fig5}
\end{center}
\end{figure}

The results of our analysis are presented in Fig.~\ref{Fig5} for
the same kinematics as in Fig.~\ref{Fig4}.  Again, we compare the
scaling functions $f_L$ and $f_T$ for electrons with those of
neutrinos and antineutrinos. Experimental $(e,e')$ data are also
included for reference. Top and bottom panels refer to different
assumptions concerning the electromagnetic form factors entering
in $(e,e')$ reactions. First, the curves $f_L$ and $f_T$ in the
top panel have been obtained assuming $G_M^n=-G_M^p$, i.e., the
proton and neutron magnetic form factors are simply set equal in
size and opposite in sign.  In this way, we remove the isoscalar
contribution in the magnetic form factor which, consequently,
becomes purely isovector. The proton and neutron electric form
factors are not modified. Therefore, results for $(e,e')$ in the
top panel of Fig.~\ref{Fig5} reflect the scaling functions where
the isoscalar contribution only enters through the electric
content of the nucleons. It is important to point out that the
strength of the transverse nuclear response function $R^T$
increases significantly when isoscalar contributions in the
magnetic form factor are removed.  However, concerning the scaling
functions, the results in Fig.~\ref{Fig5}, when compared with
Fig.~\ref{Fig4}, show that the discrepancy between $f_L$ and $f_T$
gets smaller because a visible decrease occurs for $f_T$. In other
words, removing the isoscalar contribution in $G_M$ leads to a
weaker violation of the zeroth-kind scaling property (within the
RMF context).

In the bottom panel of Fig.~\ref{Fig5} we show the results
corresponding to no convective terms, i.e., the electric form
factors for protons and neutrons (in the electromagnetic sector) are
forced to be zero. This is a very drastic assumption which leads to
nuclear electromagnetic longitudinal responses $R^L$ being very
close to zero, having only relativistic-order contributions
involving $G_{Mp}$ or $G_{Mn}$. The reason to consider this
non-convective limit comes from the effects introduced by the
isoscalar/isovector contributions in the electric form factors of
the nucleon. Obviously, the neglect of convective terms yields
neither isoscalar nor isovector contributions. While being aware of
the important differences introduced in the cross sections due to
the assumption $G_E^p=G_E^n=0$, the analysis of scaling functions,
constructed by taking the proper ratios between the nuclear
responses and the single-nucleon ones, requires the use of the same
approach (no isoscalar or no convective terms) in both the numerator
(hadronic dependence) and denominator (single-nucleon). Hence, it is
instructive to explore the behavior of the scaling functions in such
approximations. The results in the bottom panel of Fig.~\ref{Fig5}
show that a unique ({\sl universal}) scaling function emerges from
the analysis of $(e,e')$ calculated cross sections.  Moreover, this
function (without isoscalar terms) agrees with the one evaluated
from $(\nu_\mu,\mu^-)$ and $(\overline{\nu}_\mu,\mu^+)$ processes
and with $f_{exp}(\psi')$ extracted from the analysis of
longitudinal $(e,e')$ world data.

In conclusion, we have investigated isospin effects in
quasielastic electron and CC neutrino-nucleus cross sections
making use of the RIA and various descriptions of FSI with special
emphasis placed on the RMF approach. Our main results can be
summarized as follows:
\begin{itemize}
\item
  The RMF description of FSI leads to a clear violation of zeroth-kind
  scaling. This violation only occurs in this model,
  and not in other approaches based on non-relativistic, semi-relativistic
  or relativistic plane-wave models.
\item
  The breaking of zeroth-kind scaling has been proven to be due to the
  important dynamical enhancement of the lower components in the Dirac
  wave functions (mainly in the final state) produced by the
  use of strong scalar and vector relativistic potentials.
\item
  The balance between proton and neutron contributions in the scaling
  functions, namely third-kind scaling, evaluated from $(e,e')$ reactions
  is also significantly
  affected by dynamical relativistic effects.  This contrasts with NR,
  SR and RPWIA approaches, where protons and neutrons lead to similar
  scaling functions. The same comment applies to the effective
  momentum approach.
\item
  Finally, we have investigated in more depth the origin
  of the differences observed between the
  scaling functions occurring in electron scattering and CC neutrino
  reactions. Contrary to what intuition would suggest, the longitudinal scaling
function for electron scattering is found to agree with the
neutrino-scattering $f$ (which is purely transverse) much better
than does the transverse scaling function from electron scattering.
We have shown that this result is consistent with the different
roles played by isoscalar and isovector nucleon form factors in the
two processes. In part this result probably arises because the
convective effects are less important for the CC neutrino reaction
than for electron scattering, especially for the axial-vector
contributions in the former which are dominated by spin-flip matrix
elements.
\end{itemize}
These general results complement other previous
findings~\cite{JAC06,PRL05,Amaro05} and support the essential
assumption of SuSA. Furthermore, the zeroth-kind scaling violations,
not present in other models, may give us some clues as to how to
proceed when trying to disentangle the separate roles played by
isoscalar and isovector form factors. Obviously, more precision
data, particularly for separated longitudinal and transverse
contributions to the cross sections and their effects on the scaling
functions, are needed. The present experimental results indicate an
excess of transverse strength below pion production threshold. It is
important to identify the source of this strength (FSI effects in
$L$ and $T$ channels, relativistic dynamics, effects beyond the
impulse approximation...).  The present study should be considered
as a step in this direction.

\section*{Acknowledgements}
This work was partially supported by DGI (Spain) and FEDER funds
under Contracts Nos.  FIS2005-01105, FIS2005-00810, FPA2006-13807
and FPA2006-07393, by the Junta de Andaluc\'{\i}a, by the Comunidad
de Madrid and UCM (project 910059 `Grupo de F\'{\i}sica Nuclear'),
and by the INFN-CICYT collaboration agreement (project ``Study of
relativistic dynamics in electron and neutrino scattering'').  It
was also supported in part (TWD) by the U.S. Department of Energy
Office of Nuclear Physics under contract No. DE-FG02-94ER40818. Part
of the computations were performed in the `high capacity
computational cluster for physics' of UCM, funded in part by UCM and
in part with FEDER funds.


\end{document}